\documentstyle{mn}
\title[VLA 8.4-GHz monitoring of the lens system B1933+503]{VLA 8.4-GHz 
monitoring observations of the CLASS gravitational lens B1933+503}
\author[A.~D.~Biggs et al.]{A.~D.~Biggs,$^1$ E.~Xanthopoulos,$^1$ 
I.~W.~A.~Browne,$^1$ L.~V.~E.~Koopmans$^2$ and
\newauthor C.~D.~Fassnacht$^3$\\ 
$^1$University of Manchester, Jodrell Bank Observatory, Macclesfield, 
Cheshire SK11 9DL\\ 
$^2$University of Groningen, Kapteyn Astronomical Institute, Postbus 800, 
9700 AV Groningen, the Netherlands\\ 
$^3$National Radio Astronomy Observatory, PO Box 0, Socorro, NM 87801, USA}
\date{Accepted  . 
     Received   }

\begin{document}
\maketitle
\begin{abstract}

The complex ten-component gravitational lens system B1933+503 has been 
monitored with the VLA during the period February to June 1998 with a view to
measuring the time delay between the four compact components and hence to
determine the Hubble parameter $H_0$. Here we present the results of an `A' 
configuration 8.4-GHz monitoring campaign which consists of 37 epochs with an 
average spacing of 2.8 days. The data have yielded light curves for the four 
flat-spectrum radio components (components 1, 3, 4 and 6). We observe only 
small flux density changes in the four flat-spectrum components which we do 
not believe are predominantly intrinsic to the source. Therefore the variations
do not allow us to determine the independent time delays in this system. 
However, the data do allow us to accurately determine the flux density ratios
between the four flat-spectrum components. These will prove important as
modelling constraints and could prove crucial in future monitoring observations
should these data show only a monotonic increase or decrease in the flux 
densities of the flat-spectrum components.

\end{abstract}

\begin{keywords}
galaxies: individual: B1933+503 -- cosmology: observations -- gravitational 
lensing
\end{keywords}

\section{Introduction}

B1933+503 is the most complex arcsec-scale gravitational lens system known and 
the most noteworthy system discovered in the Cosmic Lens All-Sky Survey 
(CLASS; Browne et al. 1998), a survey of approximately 15,000 flat-spectrum 
radio sources observed with the VLA at 8.4 GHz with a resolution of 200 
milliarcsec (mas). Sykes et al. (1998) report the discovery of this lens and 
present VLA and MERLIN maps that reveal up to ten components, four of which 
are compact and have flat spectra while the rest are more extended and have 
steep spectra. The ten components are believed to be the multiple images of a 
background source that consists of a flat-spectrum core (quadruply imaged) and 
two compact lobes symmetrically situated on opposite sides of the core (one 
quadruply imaged and the other doubly imaged). A {\em Hubble Space Telescope
(HST)}/WFPC2 image ($I$-band) of B1933+503 (Sykes et al. 1998) shows a faint 
galaxy with a compact core (the lensing galaxy) but none of the images of the 
background object are detected. Spectra taken with the Keck telescope have 
given a redshift for the lensing galaxy of 0.755 (Sykes et al. 1998). Sub-mm 
observations of B1933+503 (Chapman et al. 1998) pointed to the fact that the 
source of the lens system should be lying above a redshift of 2 and that has 
been confirmed recently by observations made with the United Kingdom Infra-red
Telescope (UKIRT) that have yielded a redshift of 2.62 for the background 
source (Norbury et al., in preparation).

{\em HST} NICMOS F160W-band observations (Marlow et al. 1998) of B1933+503 
have uncovered the infra-red counterparts to two of the four compact 
components. These same two components were also detected in a VLBA 1.7-GHz map 
with a resolution of 6 mas. Marlow et al. suggest that the failure to detect 
the other two flat-spectrum components in the infra-red can be due to either 
rapid variability on a time-scale less than the time delay or due to 
extinction in the interstellar medium (ISM) of the lensing galaxy. The absence 
of these same two components in the radio could arise due to scatter-broadening
of the images when propagating through the ISM of the lens galaxy as well as 
from time variability.

B1933+503 offers a unique opportunity to construct an accurate mass model due 
to the wealth of observational constraints that are available. As well as the 
large number of images, yielding both positional and flux density ratio 
constraints, some of these are extended. Perhaps more crucial is that three 
epochs of VLA 15-GHz data (Sykes et al. 1998) suggest that the flat-spectrum 
components are variable, by as much as 33~per~cent over the period 
July to September 1995. This means that it should be possible to determine the 
relative time delays from flux density monitoring. Taken together this makes 
B1933+503 a promising system for a determination of the Hubble constant, $H_0$ 
(Refsdal 1964). The observational constraints given by Sykes et al. (1998) 
have been used as a starting point for the model presented by Nair (1998). The 
time delays of the flat-spectrum components with respect to component 1 
derived from this model are approximately 8~d for component 3, 7~d for 
component 4 and 9~d for component 6.

In this paper we present the results of a four-month monitoring campaign using
the VLA in its `A' configuration at a frequency of 8.4~GHz. This frequency
was chosen over 15~GHz (where the magnitude of flux density variations are 
likely to be greater) for a number of reasons. As well as the sensitivity of
the VLA being greater at 8.4~GHz, the phase stability of the array is much
poorer at 15~GHz. In addition, VLA 15-GHz monitoring of the lens system 
B0218+357 (Biggs et al. 1999) showed that at this frequency the data are
subject to calibration errors of a few per~cent due to the gains of the
antennas varying with elevation. In Section~2 we describe the 
observations and observing strategy as well as the reduction methods that we 
used on the data. In Section~3 we show the light curves that were obtained 
from the data and we follow this with a discussion of the results.  

\section{Observations and Data Reduction}

B1933+503 was observed between February 16 1998 and June 1 1998, during which 
time the VLA was in its `A' configuration. The 37 epochs were separated, on 
average, by 2.8~d. The observations were made simultaneously in two frequency 
bands (IFs) each with a bandwidth of 50~MHz and centre frequencies of 8.4351 
and 8.4851~GHz respectively. This gives an angular resolution of approximately 
200 mas.  

\begin{figure}
\begin{center}
\setlength{\unitlength}{1cm}
\begin{picture}(5,8.5)
\put(-1.6,-0.6){\includegraphics{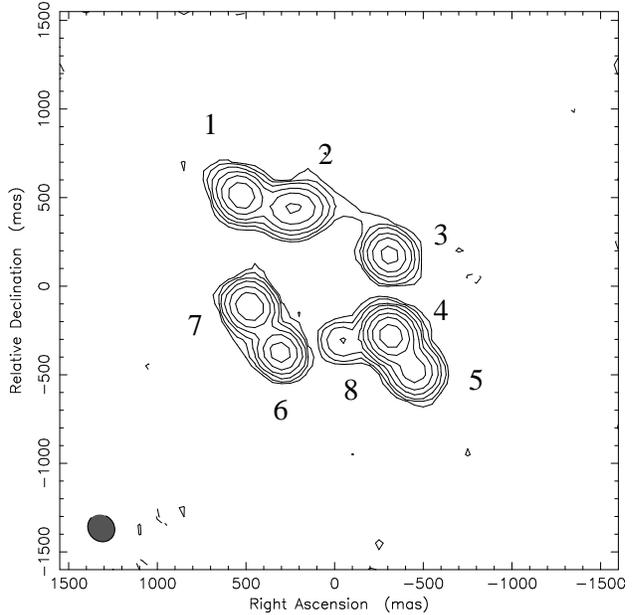}}
\end{picture}
\caption{The VLA 8.4-GHz map of B1933+503 made from ten epochs of monitoring
data. The map center is at RA 19 34 30.932 and DEC +50 25 23.502 (J2000 
coordinates). The contours are equal to $-0.5, 0.5, 1, 2, 4, 8, 16, 32, 64$ 
per cent of the maximum value of 14.4~mJy~beam$^{-1}$.}
\label{1933map}
\end{center}
\end{figure}

The observing strategy was kept as consistent as possible over the period of 
the monitoring. At each epoch the B1933+503 observations were split into two 
or three scans, each of approximately 6 min duration. Before and after each of 
the B1933+503 scans were separate ($\sim$one min duration) scans of the 
steep-spectrum source B1943+546 (Patnaik et al. 1992) which was used as the 
phase and amplitude calibrator. B1943+546 has all the requirements of a good
primary calibrator being close to the target, having a strong correlated flux 
density and being relatively unresolved with the VLA at this frequency. The 
duration of the combined observations of both sources at a typical epoch was 
30 min, but several epochs consisted of 20 min observing time. 

\begin{figure*}
\begin{center}
\setlength{\unitlength}{1cm}
\begin{picture}(5,12)
\put(-6,13){\includegraphics{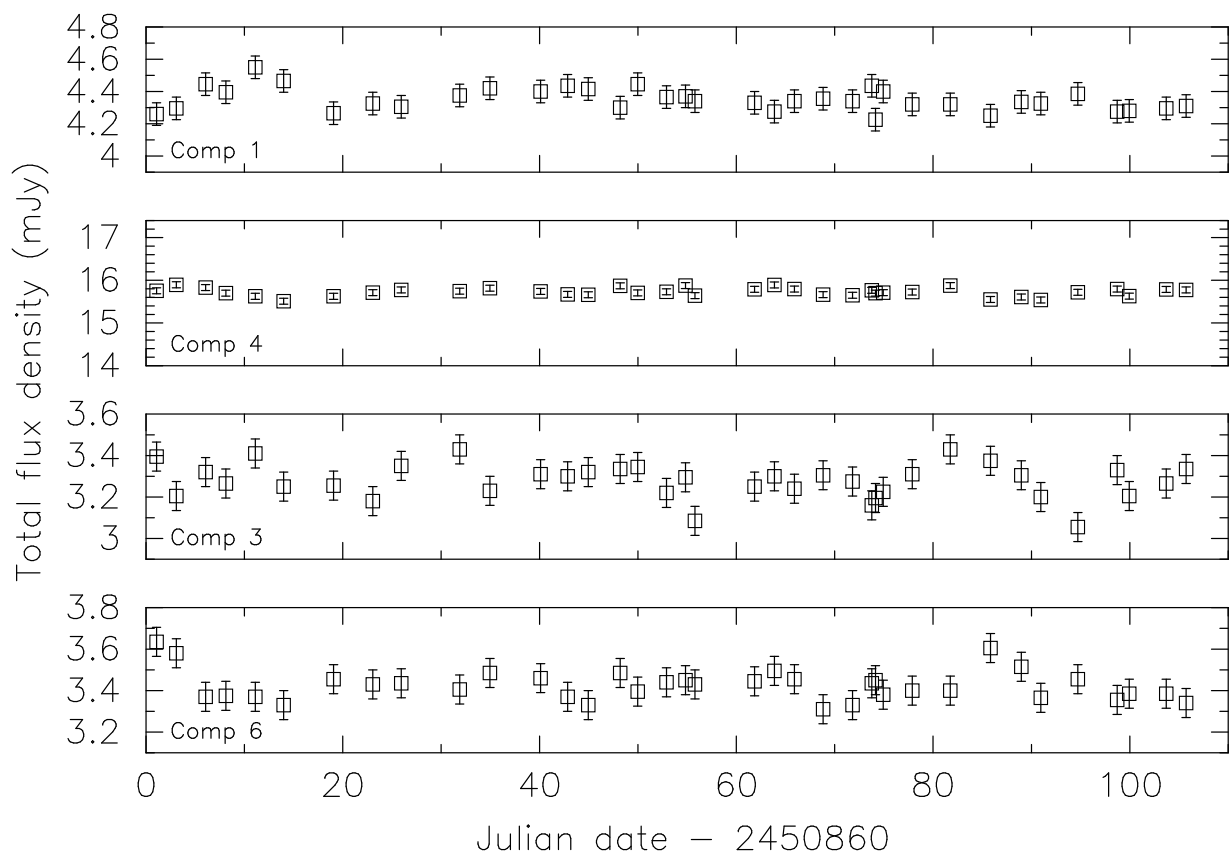}}
\end{picture}
\caption{The 8.4~GHZ total flux density light curves for the four 
flat-spectrum components of B1933+503. The components have been plotted in the 
order in which they are predicted to vary, beginning with component 1 at the 
top of the figure and followed by components 4, 3 and 6 respectively. The 
range along the y-axis has been set to a fixed fraction ($\sim20\%$) of the 
average flux density for each component.}
\label{fluxfig}
\end{center}
\end{figure*}

Calibration was performed using the NRAO Astronomical Image Processing 
Software package {\sc aips} and each IF was calibrated separately. The first 
step was to make the best possible map of 1943+546 as this calibrator is 
slightly resolved at the observing frequency. This was done by performing a 
first order calibration of several best epochs of 1943+546 data assuming the
flux density of 610~mJy measured by Patnaik et al. (1992). This source has a 
steep spectrum and is found from VLBI maps (Polatidis et al. 1995; Xu et al. 
1995) to be dominated by emission from two extended lobes on scales of 
$\sim$50~pc. Therefore it is unlikely that the source's flux density will have 
changed significantly since this measurement or that it will have changed over 
the period of the monitoring. These epochs were then combined and iteratively 
mapped and self-calibrated. The structure found from this map agrees well with 
that seen in the above mentioned VLBI maps. The CLEAN components from the map 
were then used as a model input to the {\sc aips} task {\sc calib} to derive 
telescope gain solutions at each epoch of observation. These gain solutions 
were then applied to the target B1933+503.

The positions and approximate flux densities of each component of B1933+503
were first established by making a map of the source. In order to make a high 
dynamic-range map of the source we combined ten epochs of observations with 
the {\sc dbcon} task. As the images of the radio core are seen in higher 
resolution MERLIN maps to be compact, a traditional CLEAN was not used to make 
the map. Instead, a single delta component (point source) was placed at the 
approximate position of each of the compact components in the dirty map and 
the positions and flux densities of these were then optimised by model-fitting.
Once the best fit had been found, the remaining extended flux was then CLEANed 
in the usual way. The resulting map is shown in Fig.~\ref{1933map} with eight
of the ten components labelled. The remaining two components cannot be easily
distinguished due to the low resolution of the image. Higher resolution
observations and lens modelling have shown that component~2 actually consists 
of two merging images, denoted 2a and 2b by Nair (1998). The remaining
component (1a) can be barely discerned in Fig.~\ref{1933map} as a slight
extension to the north-east of component~1. This extension is much more 
obvious in a MERLIN 1.7-GHz map (Sykes et al. 1998).

The flux densities of the four compact components of B1933+503 were determined 
by model-fitting directly to the $(u,v)$ data, and to each IF individually,
using the {\sc difmap} (Pearson et al. 1994; Shepherd 1997) model fitter. Each 
epoch of monitoring data was fit using the model derived from the map with all 
flux densities and positions held constant, except for the flux densities of 
the four compact components which were allowed to float. The model fit was 
interspersed with two iterations of phase self-calibration followed by a final 
overall amplitude correction i.e. one correction for each antenna for the 
entire observing period, normalising against the extended structure whose 
brightness should not vary over the monitoring observations. These amplitude
corrections are never more than a few per cent per telescope. The reduced 
$\chi^2$ indicated a reasonably good fit to each epoch, with a mean value of 
1.35 and an rms scatter of 0.04 for each IF's data. At this point the two flux 
density measurements at each epoch from each IF are averaged together.

\begin{table*}
\begin{center}
\caption{Flux densities of the flat-spectrum radio components at 8.4~GHz for 
three different epochs (from 1994, 1995 and 1998). Typical flux density errors 
are 0.4~mJy. Flux density ratios of components 3, 4 and 6 with respect to 1.}
\begin{tabular}{lllllll}
Component & \multicolumn{3}{c}{Flux density (mJy)} & \multicolumn{3}{c}{Flux density ratio} \\
    & \multicolumn{3}{c}{8.4 GHz}  & \multicolumn{3}{c}{8.4 GHz} \\
    & (1994)  &  (1995)  &  (1998) & (1994)  &  (1995)  &  (1998)\\
1 & 4.6  & 4.3  & 4.3  & 1.00 & 1.00 & 1.00\\
3 & 3.5  & 3.4  & 3.3  & 0.76 & 0.79 & 0.77\\
4 & 17.6 & 15.7 & 15.7 & 3.83 & 3.65 & 3.65\\
6 & 3.8  & 3.6  & 3.4  & 0.83 & 0.83 & 0.79\\
\end{tabular}
\label{fluxtab}
\end{center}
\end{table*}

\section{Results}

The averaged total flux density light curves for components 1, 4, 3 and 6 are 
shown in Fig.~\ref{fluxfig} and are plotted from the top of the figure 
downwards in the order in which they are expected to vary. The range 
encompassed by the y-axis is, for each component, equal to the same fraction 
($\sim20\%$) of the average flux density calculated over the period of the 
monitoring. The error bars attached to each data point have been estimated 
from a consideration of the difference between the flux densities derived from 
each IF at each epoch. For each component, a frequency histogram of the flux 
density difference at each epoch is approximately Gaussian with a mean that 
does not differ from zero by more than 0.02~mJy. The 1$\sigma$ error derived 
using this method for each component is very similar and has been fixed at 
0.07~mJy in Figure~\ref{fluxfig}. This is in fact only a lower limit to the 
uncertainty in the flux densities as it does not take into account errors that 
are correlated in each IF. These include errors arising from overall amplitude 
miscalibration and those that are a function of the $(u,v)$ coverage of the 
data which is essentially the same for each IF of a given epoch.

From the variations seen in Figure~\ref{fluxfig} it does not appear that the
source has varied significantly during the course of our monitoring campaign. 
This conclusion is supported in a number of ways. First, the mean flux 
density and rms percentage scatter about this mean for components 1, 4, 3 and 
6 are: 4.35~mJy, 15.73~mJy, 3.25~mJy and 3.43~mJy, and 1.6, 0.6, 2.6, and 2.2
per cent respectively. What is apparent is that the rms scatter in the light 
curves is inversely proportional to the mean flux density of the component. If 
there were actually a dominant signal of intrinsic variability then it would 
be expected that the fractional scatter in the data would be equal for each 
component. This is not the case. This can be seen most clearly by comparing 
the `variability' seen in the brightest component (4) which is much less than 
that seen in the other three fainter components. Also, a preliminary time 
delay analysis using both a chi-squared minimisation and the Dispersion method 
of Pelt et al. (1996) result in best-fit time delays that are totally at odds 
with those predicted by lens models of this system. The rms scatters quoted 
above are in general larger than expected from thermal noise considerations 
alone (by factors of $\sim$2--3) and indicate that the data are corrupted by 
errors stemming from deconvolution and model fitting. Errors introduced by
incorrect amplitude calibration from epoch to epoch must be present at a low
level ($\leq0.6\%$) as these would produce the same fractional scatter in each 
component.

Table~\ref{fluxtab} lists the flux densities of the four flat-spectrum 
components measured from the map in Fig.~\ref{1933map} together with previous 
values obtained at 8.4~GHz with the VLA in `A' configuration on July 6, 1995 
(Sykes et al. 1998) as well as those determined from the original VLA snapshot 
observation taken in 1994. We see that although there were no changes between 
the two most recent epochs there are signs that the overall flux density of 
these compact components in 1994 was higher than that in the last two epochs. 
As flux density variability has been reported at 15~GHz by Sykes et al. (1998) 
and because a comparison of the MERLIN and VLA maps by the same authors 
support variations in the relative flux densities of the same components we 
think that this source has been, and will probably again be, variable.

\section{Concluding remarks}

We have presented the results of a four month program of monitoring the 
ten-component lens system B1933+503 with the VLA which have not allowed us to
measure the independent time delays in this otherwise promising system for
determining $H_0$. The component light curves show fluctuations of 0.7--2.5 
per cent which we do not believe represent source variability. Instead they 
are more likely to result from a combination of thermal noise, deconvolution 
and model fitting errors. Since the system has already shown indications of 
variability (Sykes et al. 1998), we think that we have caught this system in 
a quiescent phase. For this reason, we have embarked on a second monitoring 
campaign that commenced in June 1999, also with the VLA at 8.4~GHz. An 
advantage that we have now over the previous attempt is that the flux density
ratios between the components are known to high accuracy. Since we did not 
detect any variability during the 4 month monitoring campaign, which is a 
period significantly longer than the predicted time delay of 7--9~d, we have  
determined the intrinsic flux density ratios (Table~\ref{fluxtab}). Therefore, 
should this second set of observations again not show reliable features that 
can be identified in each light curve, but do show a monotonic increase or 
decrease in flux density, a time delay may still be determined. This has 
previously been demonstrated by Koopmans et al. (1999) for the CLASS lens 
B1600+434.  

\section*{Acknowledgments}
This research was supported by European Commission, TMR Programme, Research 
Network Contract ERBFMRXCT96-0034 ``CERES''. The VLA is operated by the 
National Radio Astronomy Observatory which is supported by the National 
Science Foundation operated under cooperative agreement by Associated 
Universities, Inc. ADB acknowledges the receipt of a PPARC studentship.

{}
\end{document}